\documentclass [12pt] {article}
\usepackage{graphicx}
\begin {document}
\baselineskip=30pt
\parskip 10pt plus 1pt
\vskip 2pc
\begin {center}
 {\Large{\bf Quantum dynamics of the dissipative
two-state system coupled with a sub-Ohmic bath }} \centerline {\bf
 Zhiguo L\"u, Hang Zheng} \centerline{\bf Department of
Physics, Shanghai Jiao Tong University, } \centerline{\bf Shanghai
200240, China}
\end {center}
\begin {center}
\begin {minipage} {5.5 in}

\baselineskip 18pt

\textbf{Abstract}. The decoherence of a two-state system coupled
with a sub-Ohmic bath is investigated theoretically by means of the
perturbation approach based on a unitary transformation.  It is
shown that the decoherence depends strongly and sensitively on the
structure of environment. Nonadiabatic effect is treated through the
introduction of a function $\xi_k$ which depends on the boson
frequency and renormalized tunneling.  The results are as follows:
(1) the non-equilibrium correlation function $P(t)$, the dynamical
susceptibility $\chi''(\omega)$ and the equilibrium correlation
function $C(t)$ are analytically obtained for $s\leq 1$; (2) the
phase diagram of thermodynamic transition shows the
delocalized-localized transition point $\alpha_l$ which agrees with
exact results and numerical data from the Numerical Renormalization
Group; (3) the dynamical transition point $\alpha_c$ between
coherent and incoherent phase is explicitly given for the first
time. A crossover from the coherent oscillation to incoherent
relaxation appears with increasing coupling (for $\alpha > \alpha_c
$, the coherent dynamics disappear); (4) the Shiba's relation and
sum rule are exactly satisfied when $\alpha \leq \alpha_c $; (5) an
underdamping-overdamping transition point $\alpha_c^{*}$  exists in
the function $S(\omega)$.  Consequently, the dynamical phase
diagrams in both ohmic and sub-Ohmic case are mapped out. For
$\Delta \ll \omega_c$, the critical couplings ($\alpha_l, \alpha_c$
and $\alpha_c^{*}$) are proportional to $\Delta^{1-s}$.


\end{minipage}
\end{center}
E-mail: zglv@sina.com and hzheng@sjtu.edu.cn\vfill \eject
\parindent 1.5pc \vskip 0.5pc

\section{Introduction}

 A two-state system (TSS) coupled to a dissipative
environment offers a unique testing ground for exploring
fundamental physics of quantum mechanics behaviors such as
tunneling and decoherence\cite{rmp,book}. It also provides a
paradigmatic standard model for studying quantum
computation\cite{qc}.  An interesting problem in this type of
system is decoherence resulting from the influence of an
environment, which is usually represented by an infinite set of
harmonic oscillators. This system can be described by the
spin-boson model (SBM), which reads
\begin{eqnarray}
&&H=-{\frac{1}{2}}\Delta \sigma _{x}+\frac{1}{2}\epsilon\sigma_{z}
+\sum_{k}\omega _{k}b_{k}^{\dag }b_{k}
+{\frac{1}{2}}\sum_{k}g_{k}(b_{k}^{\dag }+b_{k})\sigma _{z},
\end{eqnarray}
Here $\sigma_{x}$ and $\sigma_{z}$ are the usual Pauli spin
matrices, $\Delta$ is the bare tunneling matrix element between
the two states, $\epsilon$ describes the bias of the system.
Throughout this paper we set $\hbar =1$ and $k_{B}=1$. The
quantity $g_{k}$ represents the coupling strength of TSS to the
kth oscillator. Without dissipative environment, the TSS exhibits
coherent tunneling between the two states, while the coupling to
the environment will result in the lose of its phase coherence.

Impossible as an exact solution of SBM is, both the equilibrium
and non-equilibrium dynamics in this model have been widely
studied by using the Path-integral formalism, variational method,
real time quantum Monte Carlo, numerical renormalization group
(NRG), flow equations, etc[1-26].  Much of the physics underlying
the model has not been revealed by numerical methods, such as the
coherent-incoherent transition\cite{Bulla,sm}. The main interest
is to understand how the quantum environment influences the
dynamics of TSS, in particular, how dissipation destroys quantum
coherence. On the one hand, when the system can be prepared in one
of the two states by applying a strong bias for times $t<0$ and
then let it evolve for $t>0$ in zero bias, the non-equilibrium
correlation function $P(t)$ is of primary interest\cite{rmp,gui}.
Moreover, $P(t)$ can be directly measured by the technique of
muon-spin rotation\cite{book2}. On the other hand, while the
initial state preparation is not realizable, the interest lies in
the equilibrium correlation function $C(t)$ related to the
cross-section of neutron scattering and the susceptibility
$\chi(\omega)$\cite{rmp}.

The effect of a harmonic environment is characterized by a
spectral density
$J(\omega)=2\alpha\omega_{s}^{1-s}\omega^s\theta(\omega_{c}-\omega)$
with the dimensionless coupling strength $\alpha$, the upper
cutoff $\omega_{c}$ and the step function $\theta(x)$.  An
additional energy scale $\omega_{s}$ is introduced, but only the
combination $\alpha \omega_{s}^{1-s}$ has fundamental
significance. In this paper we assume that the high energy cutoff
$\omega_{c}$ is much larger than all other scales. It is
convenient to set $\omega_s = \omega_c/100$ in this paper except
the special notation. The index $s$ accounts for various physical
situations around the TSS. For example, in solid materials where
acoustic phonon provides the most efficient damping mechanism,
according to the Debye model, $s=3$, which belongs to a
super-Ohmic bath($s>1$). $s=1$ and $s<1$ stand for Ohmic bath and
sub-Ohmic one, respectively. There are suggestions to model $1/f$
noise found in experiments by a limiting sub-Ohmic
bath\cite{naka,pala,shnir}. In terms of the renormalization group,
sub-Ohmic coupling represents a relevant perturbation\cite{bach}.
Moreover, there are various claims that the particle is always
localized in the sub-Ohmic case for zero temperature based on
noninteracting blip approximation (NIBA)\cite{rmp,book}.  As was
pointed out in Ref. \cite{1.}, the NIBA is unreliable for studying
the long-time behaviour of equilibrium correlation functions since
it gives an exponential decay instead of an algebraic decay.
Another tool for treating spin-boson problem, adiabatic
renormalization, is invalid in the sub-Ohmic case\cite{1.}. Due to
technical difficulties, little result about long time dynamics of
$s<1$ at zero temperature is known.

The study of SBM with a sub-Ohmic bath for zero bias case has
attracted much attention recently [5-12]. The flow equation method
has been applied by S. Kehrein, A. Mielke and T. Stauber to this
model\cite{1.,2.,3.}. Some important properties were predicted, such
as a transition from delocalization for weak coupling to
localization for strong coupling and the calculation on the
equilibrium spectral function $C(\omega)$. However, as was pointed
out by Stauber, this approach did not yield a correct normalization
condition nor a satisfying result about the Shiba relation for a
certain parameter regime\cite{3.}. This approach seems difficult to
calculate the quantum dynamics of SBM, especially the
non-equilibrium correlation function $P(t)$. Besides, the NRG
method, a powerful numerical tool, was employed by Bulla, Tong,
Vojta, et.al to investigate the sub-Ohmic case and provide some
reliable results for both static and dynamic quantities in the whole
range of model parameters and
temperatures\cite{Bulla,Tong,vojta,anders}. They focused on the
quantum critical behavior from a delocalized state to a localized
one and found that the Shiba relation was fulfilled within an error
of about $10\%$ in the Ohmic case, but did not provide a check in
the sub-Ohmic case\cite{Bulla}. Very recently, the NRG study for
sub-Ohmic case shows that the dynamical properties of the
delocalized phase are not dominated only by a single energy scale
$\Delta_r$\cite{anders}. In their treatment, the energy scale
$\omega_s$ was set equal to the high energy cutoff $\omega_c$. To
the best of our knowledge, the dynamics transition between coherent
oscillation and incoherent relaxation has not been given explicitly
so far.  Recently, Chin and Turlakov have used the variational
method originally proposed by Silbey and Harris for the sub-Ohmic
bath\cite{chin}. They also paid more attention to the transition
between delocalized phase(the effective tunneling $\Delta_r>0$) and
localized one ( $\Delta_r=0$) at both $T=0$ and $T>0$. Even if they
pointed out that dynamical and thermodynamics criteria for the
transition should be expected different and sensitive to
non-adiabatic mode. In their discussion, they have ignored the
dynamical effect of the perturbation. Thus, it is difficult for the
variational ansatz to make detailed statement about quantum dynamics
beyond Born-Markov approximation. Moreover, they worked with
$\omega_s \neq \omega_c$ and $\omega_c \rightarrow \infty $. As far
as we know, there is no numerical and analytical approach that can
give the dynamical transition of nonequilibrium correlation
function, exact Shiba's relation and sum rule in the sub-Ohmic case.
In the present paper, we extend a unitary transformation proposed by
one of our authors to investigate quantum dynamics of the sub-Ohmic
SBM at $T=0$\cite{zhe}.

The paper is organized as follows. In Sec.\textsc{II}, the
hamiltonian is separated into the unperturbed part and the
perturbed one based on a unitary transformation and perturbation
theory. In Sec.\textsc{III A}, the localization-delocalization
transition point will be clarified, compared with the results of
flow equations and NRG. The calculations of quantum dynamics
$P(t)$ and $C(t)$ will be given explicitly in Sec.\textsc{III B}.
Then, It is verified that real and imaginary parts of
$\chi(\omega)$ satisfy the Shiba's relation analytically and
numerically in Sec.\textsc{III C}. The coherent-incoherent
transition is also discussed in detail in this part. Finally,
Sec.\textsc{IV} gives discussion and further analysis of our
results and approach . Usually, people believe that perturbation
approach is not good for the dissipative SBM because of the
infrared divergence in calculating the renormalized tunneling
frequency and other physical quantities by perturbation expansion.
Here we attempt to get rid of the divergence by using a unitary
transformation.  When one take the scaling limit $\Delta
<<\omega_c$ from our obtained results, it will reproduce the known
or exact one. This approach works well for the low-temperature
region and weak coupling case with $0<\Delta<\omega_c$ far away
from the scaling limit.

\section{Unitary transformation}

In order to take into account the correlation between spin and
bosons, we present a treatment based on the unitary transformation
to $H$: $H^{\prime}= e^{S} H e^{-S}$, where the generator of the
transformation is
\begin{eqnarray}\label{unitary}
&&S=\sum_{k}\frac{g_{k}}{2\omega _{k}}\xi _{k}(b_{k}^{\dag
}-b_{k})\sigma_z.
\end{eqnarray}
A $k-$dependent function $\xi_k$ introduced in the transformation
corresponds to the displacement of each boson mode due to the
coupling to the TSS\cite{zhe,zhe2}. Its form will be determined
later by the perturbation theory.

The transformation can be done to the end and the result is
$H^{\prime }=H_{0}^{\prime }+H_{1}^{\prime }+H_{2}^{\prime }$,
where
\begin{eqnarray}
&&H_{0}^{\prime }=-{\frac{1}{2}}\eta \Delta \sigma _{x}
+\sum_{k}\omega _{k}b_{k}^{\dag }b_{k}
-\sum_{k}\frac{g_{k}^{2}}{4\omega _{k}}\xi _{k}(2-\xi _{k}), \\
&&\eta =\exp [-\sum_{k}\frac{g_{k}^{2}}{2\omega _{k}^{2}}\xi _{k}^{2}], \\
&&H_{1}^{\prime }={\frac{1}{2}}\sum_{k}g_{k}(1-\xi
_{k})(b_{k}^{\dag
}+b_{k})\sigma _{z} -{\frac{1}{2}}\eta \Delta i\sigma_y\sum_{k}%
\frac{g_{k}}{\omega _{k}} \xi _{k}(b_{k}^{\dag }-b_{k}), \\
&&H_{2}^{\prime }=-{\frac{1}{2}}\Delta\sigma_x\left( \cosh \{\sum_{k}\frac{%
g_{k}}{\omega _{k}}\xi _{k}(b_{k}^{\dag }-b_{k})\}-\eta \right)  \nonumber \\
&&-{\frac{1}{2}}\Delta i\sigma_y\left( \sinh
\{\sum_{k}\frac{g_{k}}{ \omega
_{k}}\xi _{k}(b_{k}^{\dag }-b_{k})\}-\eta \sum_{k}\frac{g_{k}}{ \omega _{k}}%
\xi _{k}(b_{k}^{\dag }-b_{k})\right)
\end{eqnarray}
$H^{\prime}_0$ is the unperturbed part of $H^{\prime}$. Obviously,
since the spin and bosons are decoupled in this part,
$H_{0}^{\prime}$ can be solved exactly. The eigenstate of
$H_{0}^{\prime }$ is a direct product: $|s\rangle
|\{n_{k}\}\rangle$, where $|s\rangle $ is the eigenstate of
$\sigma_x$: $ |s_{1}\rangle =\left(
\begin{array}{c}
1 \\
1
\end{array}
\right) $ or $|s_{2}\rangle =\left(
\begin{array}{c}
1 \\
-1
\end{array}
\right) $, and $|\{n_{k}\}\rangle$ is the eigenstate of bosons
with $n_{k}$ bosons for mode $k$. In particular,
$|\{0_{k}\}\rangle$ is the vacuum state in which $n_{k}=0$ for
every $k$. The ground state of $H_{0}^{\prime } $ is
\begin{equation}
\ |g_{0}\rangle =|s_{1}\rangle |\{0_{k}\}\rangle .
\end{equation}

$H_{1}^{\prime }$(first-order terms) and $H_{2}^{\prime }$
(including second order terms and higher ones )are treated as
perturbation and they should be as small as possible. For this
purpose $\xi _{k}$ is determined as

\begin{eqnarray}
\xi _{k} &=&\frac{\omega _{k}}{\omega _{k}+\eta\Delta}.
\end{eqnarray}
Substituting this form into Eq. (5), one has
\begin{eqnarray}
&&H_{1}^{\prime
}={\frac{1}{2}}\eta\Delta\sum_{k}\frac{g_{k}}{\omega_k+\eta\Delta}[b_{k}^{\dag
}(\sigma _{z} - i\sigma_y)+ b_{k}(\sigma _{z} + i\sigma_y)].
\end{eqnarray}
It is easy to check that $H_{1}^{\prime }|g_{0}\rangle =0$.  Thus,
by choosing the form of $\xi_k$, the matrix elements between the
ground and the lowest excited states are zero (we show them in the
following). It will become possible to take a perturbation
treatment based on the division of the transformed hamiltonian.
Besides, in transformed hamiltonian $H'$ $\eta\Delta$ is the
probability of diagonal transition of bosons which describes the
coherent tunnelling motion of particle. $\eta$ is determined in
Eq.(4) to make $\mbox{Tr}(\rho_B H'_2)=0$, where
$\rho_B=\exp(-\beta H_B)/\mbox{Tr}\exp(-\beta H_B)$ is the
equilibrium density operator of bosons ($H_B=\sum_k\omega_k
b^{\dag}_kb_k$). Thus one can see that the tunneling is
renormalized by this factor $\eta$ arising due to the dressing of
bosons coupled with the TSS. In other words, the particle is
surrounded by bosons cloud as it tunnels between the two states.

It is noticeable that $0\le\xi_k\le 1$, which determines the
intensity of the correlation between spin or particle presentive
for subsystem and bosons in bath: $\xi_k \sim 1$ if the boson
frequency ($\omega_k$) is much larger than the renormalized
tunneling of subsystem ($\eta\Delta$) while $\xi_k \ll 1$ for
$\omega_k \ll \eta\Delta$. Since the transformation in
Eq.(\ref{unitary}) is essentially a displacement one, physically,
one can see that high-frequency bosons ($\omega_k> \eta\Delta$)
follow the tunneling particle adiabatically(instantaneously)
because the displacement is $g_k\xi_k/\omega_k \sim g_k/\omega_k$,
which leads to a dressed particle. On the other hand,
low-frequency bosons $\omega_k < \eta\Delta$, in general, are not
always in equilibrium with the tunneling particle, and hence the
particle moves in a retarded potential arising from the
low-frequency modes. When the nonadiabatic effect dominates,
$\omega_k \ll \eta\Delta$, the displacement $g_k\xi_k/\omega_k
\approx g_k/ \eta\Delta \ll 1$, is substantially reduced.
Therefore, the effects of each boson mode on subsystem are in
nature treated separately by $\xi_k$.

The lowest excited states are $|s_2\rangle |\{0_{k}\}\rangle$ and
$|s_1\rangle |1_k \rangle$, where $|1_k \rangle$ is the number
state with $n_k=1$ but $n_{k'}=0$ for all $k' \neq k$. It is easy
to check that $\langle g_0|H'_2|g_{0}\rangle = 0$ (because of the
form of $\eta$ in Eq. (4)), $\langle\{0_{k}\}|\langle
s_2|H'_2|g_{0}\rangle =0$, $\langle 1_{k}|\langle
s_1|H'_2|g_{0}\rangle =0$, and $\langle\{0_{k}\}|\langle
s_2|H'_2|s_1\rangle |1_k \rangle=0$. Moreover, since
$H_{1}^{\prime}|g_{0}\rangle =0$, we have
$\langle\{0_{k}\}|\langle s_2|H'_1|g_{0}\rangle =0$ and $\langle
1_{k}|\langle s_1|H'_1|g_{0}\rangle =0$. Thus, we can diagonalize
the lowest excited states of $H'$ as
\begin{eqnarray}
H'=-\frac{1}{2}\eta \Delta |g_{0}\rangle\langle g_0|
     +\sum_E E |E\rangle\langle E|+\mbox{terms with higher excited states}.
\end{eqnarray}
The diagonalization is achieved through the following
transformation\cite{gui}:
\begin{eqnarray}
&&|s_2\rangle |\{0_{k}\}\rangle=\sum_E x(E)|E\rangle,\\
&&|s_1\rangle |1_k \rangle=\sum_E y_k(E)|E\rangle,\\
&&|E\rangle= x(E)|s_2\rangle |\{0_{k}\}\rangle +\sum_k
y_k(E)|s_1\rangle |1_k \rangle,
\end{eqnarray}
where
\begin{eqnarray}
&&x(E)=\left[1+\sum_k\frac{V^2_k}{(E+\eta\Delta/2-\omega_k)^2}\right]^{-1/2},\\
&&y_k(E)=\frac{V_k}{E+\eta\Delta/2-\omega_k}x(E),
\end{eqnarray}
with $V_k=\eta\Delta g_k\xi_k/\omega_k$. $E$'s are the
diagonalized excitation energy and they are solutions of the
equation
\begin{eqnarray}
&&E-\frac{\eta\Delta}{2}-\sum_k\frac{V^2_k}{E+\eta\Delta/2-\omega_k}=0.
\end{eqnarray}

\section{Quantum dynamics}
\subsection{A. Delocaliztion}

The Ohmic spin-boson model has nontrivial dynamics only for
$\alpha<1$ when $\Delta\ll\omega_{c}$. Scaling arguments, flow
equations and other methods give a renormalized tunneling,
$\Delta_r=\Delta(\frac{\Delta}{\omega_c})^\frac{\alpha}{1-\alpha}$\cite{rmp,ke2}.
When $\alpha$ goes to 1, one gets $\Delta_r=0$ which is a
transition from delocalization for weak coupling to localization
for strong coupling. However, for the sub-Ohmic bath $s<1$, there
is some confusion about the existence of such a transition within
some various approximation approaches. On the one hand, the TSS is
localized for nonzero sub-Ohmic coupling at $T=0$ which is
predicted by the NIBA\cite{rmp,book}. On the other hand, from the
mapping of SBM to Ising model, the corresponding results turn out
a transition as a function of coupling \cite{1.,sp}. Moreover, the
results of flow equations and numerical renormalization group also
support it\cite{1.,Bulla}. However, the transition point is less
clear for general $ \Delta $ and $ s $.

The renormalization of tunneling can be calculated as
\begin{eqnarray}
\eta&=& \exp\left\{-\alpha\omega^{1-s}_s\int^{\omega_c}_0
\frac{\omega^s
d\omega} {(\omega+\eta\Delta)^2}\right\}\nonumber\\
&=&\exp\left\{-\alpha\omega'^{1-s}_s\frac{\pi s}{\sin(\pi
s)}\left(\eta\Delta'\right)^{s-1}
+\alpha\omega'^{1-s}_s\sum^{\infty}_{n=1}(-1)^n\frac{n}{s-n}(\eta\Delta')^{n-1}
\right\}.
\end{eqnarray}
Here $\omega'_s \equiv\omega_s/\omega_c$ and $\Delta' \equiv
\Delta/\omega_c$. From this equation, it is self-consistent to
determine $\eta$. It is clear to see that the eigenstates of the
unperturbed part in $H^{\prime}$ are of superposition of the TSS
if $\eta
> 0$, whilst TSS becomes localized if $\eta = 0$, and no tunneling
is possible. According to the analogical treatment of the
transition condition in the work of Kohrein and Mielke\cite{1.},
the critical transition condition from delocaliztion to
localization can be derived.
\begin{eqnarray}
\eta=\exp{\left\{-\alpha\omega'^{1-s}_s\left[(\eta\Delta')^{s-1}\frac{\pi
s}{\sin\pi
s}+\sum_{n=0}^{\infty}\frac{n+1}{s-n-1}(-\eta\Delta')^n\right]\right\}}.
\end{eqnarray}
 For the scaling limit
$\Delta' \ll 1$, one gets
\begin{eqnarray}
\eta=\exp{\left\{-\alpha\omega'^{1-s}_s\left[(\eta\Delta')^{s-1}\frac{\pi
s}{\sin\pi s}+\frac{1}{s-1}\right]\right\}}
\end{eqnarray}
If the condition
\begin{eqnarray} \label{delocal}
\alpha\omega'^{1-s}_s\frac{\pi s(1-s)}{\sin(\pi
s)\Delta'^{1-s}}\leq e^{\alpha-1}
\end{eqnarray}
is satisfied, the solution for $\eta$ is finite and satisfies
\begin{eqnarray}
\eta^{1-s}\leq e^{\alpha-1}
\end{eqnarray}
otherwise $\eta=0$. It can be seen that, in the sub-Ohmic case,
there exists a quantum transition boundary separating a
delocalized phase for $\alpha<\alpha_l$ from a localized phase for
$\alpha \geq \alpha_l$. In the delocalized region, the
renormalized tunneling between the two states is finite, whereas
it is renormalized to zero in the localized phase($\eta=0$). When
$s \rightarrow 1$, from the condition (\ref{delocal}), it is easy
to get $\alpha_{l}=1$ in the scaling limit, which agrees with the
known results. Our results provide a direct evidence for a phase
transition for all $0 <s\leq 1$. Furthermore, the transition point
$\alpha_l$ is continuous as a function of $s$ from the sub-Ohmic
to Ohmic case, which is in good agreement with the NRG
results\cite{Bulla,Tong}. Compared with the transition condition
Eq. (16) in Ref.\cite{1.} obtained by flow equations, it seems
that there is a discontinuous behavior when $s$ goes below to $1$.
Moreover, no transition happens for $s>1$, in other words, the
system is always delocalized in the super-Ohmic case. As is shown
in the condition (\ref{delocal}), the critical coupling $\alpha_l$
follows a power law as a function of the bare tunneling, $\alpha_l
\propto \Delta^{1-s}$, which is in good agreement with the
conjecture of NRG\cite{Tong}. From above discussion, it indicates
that the transition condition is sensitive to the index $s$ in the
sub-Ohmic case.

$\eta$ as a function of the dimensionless coupling strength
$\alpha (\frac{\omega_s}{\Delta})^{1-s}$ is shown in Fig. 1. There
is a discontinuous jump from a finite value to $\eta=0$ as $\alpha
\rightarrow \alpha_l$ for $0<s<1$, namely, it is a first-order
transition which is different from the NRG result\cite{Bulla,
Tong}. However, for $s=1$, $\eta$ can continuously change from one
to zero as $\alpha$ increases from zero to $\alpha_l \sim 1$. For
$\alpha>1$ in the scaling limit, the delocalized transition occurs
which agrees with the known result in literature.

Our approach has given a clear evidence for the transition from a
delocalized phase to a localized one in the sub-Ohmic case. The
phase boundaries are shown in Fig. 2(a) which are determined by
the vanishing of renormalized tunneling.  NRG results are shown
for comparison. It is seen that our data agree well with those of
NRG for $s>0.6$. At the same time, there is some deviation from
our data for $s<0.5$ due to the NRG discretization
\cite{Bulla,Tong}. As displayed in Fig. 2(b), the critical
coupling $\alpha_l$ follows a power law as a function of the
tunneling, $\alpha_l \propto \Delta'^{1-s}$ for
$\Delta<<\omega_{c}$ (see Eq. \ref{delocal}), also predicted by
the NRG method\cite{Tong}.  NRG data are provided for a fair
comparison in Fig. 2(b), too. The calculated transition points are
in agreement with those obtained by the NRG method. However, our
result predicts a first-order transition, in contrast with the
second-order transition predicted by the NRG.

\subsection{B. Correlation function}

The main interest of quantum dynamics is the non-equilibrium
correlation $P(t)$ and the symmetrized equilibrium correlation
$C(t)$. When the initial state can be prepared, the evolution of
the state is of primary interest which can be described by $P(t)$.
$P(t)=\langle
b,+1|\langle+1|e^{iHt}\sigma_ze^{-iHt}|+1\rangle|b,+1\rangle$ is
defined in the Ref. \cite{rmp}, where $|+1\rangle$ is the state of
bosons adjusted to the state of $\sigma_z=+1$.

Because of the unitary transformation ($e^S\sigma_z
e^{-S}=\sigma_z$)
\begin{eqnarray}
&&P(t)=\langle \{0_{k}\}|\langle +1| e^{iH't}\sigma _{z}
e^{-iH't}|+1\rangle |\{0_{k}\}\rangle,
\end{eqnarray}
since $e^S|+1\rangle |b,+1\rangle=|+1\rangle |\{0_{k}\}\rangle$.
Using equations(10-16) the result is
\begin{eqnarray} \label{pt}
&&P(t)={1\over 2}\sum_E x^2(E)\exp[-i(E+\eta\Delta/2)t]
+{1\over 2}\sum_E x^2(E)\exp[i(E+\eta\Delta/2)t] \nonumber\\
&&=\frac{1}{4\pi i}\oint_C d\omega e^{-i\omega t}
\left(\omega-\eta\Delta-\sum_k\frac{V^2_k}{\omega+i0^+-\omega_k}\right)^{-1}\nonumber\\
&&+\frac{1}{4\pi i}\oint_C' d\omega e^{i\omega t}
\left(\omega-\eta\Delta-\sum_k\frac{V^2_k}{\omega-i0^+-\omega_k}\right)^{-1},
\end{eqnarray}
where a change of the variable $\omega=E+\eta\Delta/2$ is made.
The real and imaginary parts of $\sum_kV^2_k/(\omega\pm
i0^+-\omega_k)$ are denoted as $R(\omega)$ and
$\mp\gamma(\omega)$, respectively.
\begin{eqnarray} \label{rw1}
R(\omega)&=&\sum_{k}(\frac{\eta\Delta}{\omega_k+\eta\Delta})^{2}\frac{g_{k}^{2}}{\omega-\omega_k}\nonumber
\\
&=&-2\alpha\omega'^{1-s}_s\omega_c(\eta\Delta')^2\int_{0}^{1}\frac{x^s
dx}{(x-\omega')(x+\eta\Delta^{\prime})^2},
\end{eqnarray}
\begin{eqnarray}
&&\gamma(\omega)=2\alpha\pi\omega'^{1-s}_s\omega_{c}(\frac{\eta\Delta'}{\omega'+\eta\Delta'})^{2}
(\omega')^s\Theta(\omega_c-\omega),
\end{eqnarray}
where $\omega'\equiv \omega/\omega_c $ for simplicity. For general
$s<1$ the integration in Eq. (\ref{rw1}) should be done by Residue
Theorem,
\begin{eqnarray}
R(\omega)=-2\alpha\omega'^{1-s}_s\omega_c(\frac{\eta\Delta'}{\omega'+\eta\Delta'})^2
\left\{\sum_{n=0}^{\infty} \frac{(1-s)(-\eta
\Delta')^{n+2}+\omega' s (-\eta\Delta')^{n+1}-(\omega')^{n+2}}{n+2-s} \right. \nonumber\\
-\left.\frac{(\eta\Delta'+\omega')\eta\Delta'}{1+\eta\Delta'}-\frac{\pi}{\sin(\pi
s)}\left[(s-1)(\eta\Delta')^{s}+s(\eta\Delta')^{s-1}\omega'+(\omega')^s\cos(\pi
s)\right] \right\}.
\end{eqnarray}
The integral in Eq. (\ref{pt}) can proceed by calculating the
residue of integrand and the result is $P(t)=\cos(\omega_0
t)\exp(-\gamma t)$, where $\omega_0$ is the solution of equation
\begin{eqnarray} \label{w0}
\omega-\eta\Delta-R(\omega)=0
\end{eqnarray}
and
$\gamma=\gamma(\eta\Delta)=\alpha\pi\omega_s^{1-s}(\eta\Delta)^s/2$
(Wigner-Weisskopf approximation)\cite{scully}. The behavior of
$P(t)$ is of the form of damped oscillation. One can prove that
the solution $\omega_{0}$ of Eq. (\ref{w0}) is real for small
coupling. As the coupling $\alpha$ increases, the solution
$\omega_{0}$ becomes imaginary, thus $P(t)$ demonstrates the
incoherent dynamics. As a result, there exists a critical point
corresponding to the coherent-incoherent transition. In other
words, for the critical case, one can have $\omega_{0}=0$ and
$P(t)=\exp(-\gamma_c t)$. For $\alpha > \alpha_c$, one has $P(t)
>0$ for all times. Meanwhile, the behavior of damped oscillations
disappears and that of pure incoherence displays. When $\Delta'
\ll 1$, from Eq. \ref{w0}, one gets
\begin{eqnarray} \label{decohere}
\alpha_c = \frac{\sin(\pi s)}{2 \pi (1-s)}(\frac{\eta
\Delta'}{\omega'_s})^{1-s}
\end{eqnarray}
For $s=1$, one gets $\alpha_c=\frac{1}{2}$ which is the same as
known before. It is clear to see that $\alpha_c$ is also
proportional to $\Delta^{1-s}$ for small $\Delta$. It turns out
the sensitivity of the critical coupling to bath structure.

Fig. 3(a) shows the time evolution of the non-equilibrium
correlation $P(t)$ for the fixed coupling $\alpha
\omega'^{1-s}_s=0.1$ and $\Delta/\omega_s=10$ with different bath
types. The damped oscillation exists when $\alpha < \alpha_c$,
while it decays fast for $\alpha \sim \alpha_c$. That is to say,
the tunneling regime $\alpha < \alpha_l$, namely $\eta>0$,
consists of two qualitatively different regions, distinguished by
the presence (for $0<\alpha<\alpha_c$) or absence (for $\alpha_c<
\alpha <\alpha_l$) of tunneling oscillations in quantum dynamical
quantities\cite{Costi2}. It is predicted that there is a boundary
between the phase coherent region and pure incoherent one( the
exponential decay), which is shown in Fig. 7(a).

The behavior of $P(t)$ is shown in Fig. 3(b) for $s=0.9$ and
$\alpha=0.1$ with different tunneling $\Delta/\omega_s=1,5,10,20$,
and $30$, respectively.  The less $\Delta$ is, the faster $P(t)$
decays. It indicates that there is a nonscaling behavior for
dynamical quantities in the sub-Ohmic bath in contrast with a
scaling one in the ohmic case.

Since $e^S \sigma_z e^{-S}=\sigma_{z}$, the retarded Green's
function is
\begin{eqnarray}
G(t)&=&-i\theta (t)\left\langle [\exp (iH^{\prime }t)\sigma
_{z}\exp (-iH^{\prime }t),\sigma _{z}]_{+}\right\rangle ^{\prime
},
\end{eqnarray}
where $\langle ...\rangle ^{\prime }$ means the average with
thermodynamic probability $\exp (-\beta H^{\prime })$. The Fourier
transformation of $G(t)$ is denoted as $G(\omega)$, which
satisfies an infinite chain of equation of motion\cite{mah}. We
have made the cutoff approximation for the equation chain at the
second order of $g_{k}$ and the solution at $T=0$ is
\begin{eqnarray}
&&G(\omega)=\frac{1}{\omega -\eta\Delta -\sum_{k}V^2_k/(\omega
-\omega _{k})} +\frac{1}{\omega +\eta\Delta -\sum_{k}V^2_k/(\omega
+\omega _{k})}.
\end{eqnarray}
The equilibrium correlation function
\begin{eqnarray}
&&C(t)=\frac{1}{2}\mbox{Tr}\left\{\exp(-\beta H)[\sigma
_{z}(t)\sigma_z
+\sigma_z\sigma_z(t)]\right\} /\mbox{Tr}[\exp (-\beta H)]\nonumber\\
&&=-\frac{1}{2\pi}\int^{\infty}_{-\infty}d\omega\coth(\frac{\beta\omega}{2})
\mbox{Im}G(\omega)\exp(-i\omega t)\nonumber\\
&&=\frac{1}{\pi}\int^{\infty}_0 d\omega \frac{\gamma(\omega)}
{[\omega -\eta \Delta -R(\omega)]^2+\gamma(\omega)^2}\cos(\omega
t).
\end{eqnarray}
For general value of $\alpha\le \alpha_c$, $C(t)$ may contain both
terms of the exponential decay ones and the algebraic decay ones.
For the special value $\alpha=\alpha_c$, the exponential decay
terms disappear. In the long-time limit the first non-zero
algebraic decay term dominates which is $\sim -1/t^{s+1}$. The
time evolution of $C(t)$ is shown in Fig. 4 for $s=0.8$,
$\Delta/\omega_s=0.1$ with different couplings $\alpha=0.1, 0.3$
and $0.5$, respectively.  The time evolution of $P(t)$ is also
provided to a fair comparison in this figure. It is remarkable
that there are distinguishable characters between $C(t)$ and
$P(t)$ for larger coupling, while the difference between them is
not apparent for small coupling. This is an indication that $P(t)$
and $C(t)$ possess a similar structure for small coupling at
intermediate times. However, they differ distinctively at long
time limit.

In Fig. 5 we show $C(\omega)$ for $s=0.6$ (Fig. 5a)and
$s=0.3$(Fig. 5b).  For weak coupling, there is  a peak near the
renormalized tunnelling $\Delta_r$. However, with increasing
coupling, it is observed that part of the spectral weight has
transferred to lower frequencies with a shoulder feature. It
indicates that the dynamical properties in the sub-Ohmic case is
not determined only by a single energy scale $\Delta_r$ which
agree with the conclusion of the NRG \cite{anders}

\subsection{C. Shiba's relation and sum rule}

The susceptibility $\chi(\omega)=-G(\omega)$, and its imaginary
part is
\begin{eqnarray}
&&\chi^{\prime\prime}(\omega)= \frac{\gamma(\omega)\theta(\omega)}
{[\omega -\eta \Delta -R(\omega)]^2+\gamma^2(\omega)}
+\frac{\gamma(-\omega)\theta(-\omega)} {[\omega +\eta \Delta
+R(-\omega)]^2+\gamma^2(-\omega)}.
\end{eqnarray}

The $\omega \to 0$ limit of
$S(\omega)=\chi^{\prime\prime}(\omega)/J(\omega)$ is
\begin{eqnarray}
&&\lim_{\omega\to 0}\frac{\chi^{\prime\prime}(\omega)}{J(\omega)}
=\frac{\pi} {(\eta \Delta+R(\omega=0))^2}.
\end{eqnarray}
With a Kramers-Kronig relation and a fluctuation-dissipation
theorem,  the static susceptibility $\chi_0$ can be directly
extracted
\begin{eqnarray}
&&\chi^{\prime}(\omega=0)=\frac{1}{2}\int_0^{\infty} d
\omega\frac{C(\omega)}{\omega} .
\end{eqnarray}
Thus, there exist an important relation connecting the zero
frequency behavior of the spectral function to static
susceptibility, which is known as the Shiba's
relation\cite{vol,Keil,Costi,sas,shiba}
\begin{equation}
\lim_{\omega\to 0}\frac{C(\omega)}{(2 \chi_0 )^2 J(\omega)}= 1.
\end{equation}
It constitutes an important check of our approach.

Approximation schemes such as NIBA and numerical approaches based
on Monte Carlo simulations can not be used to verify the Shiba's
relation since they failed to predict the long time
behavior\cite{2.}. Results from both real time renormalization
group and NRG in the Ohmic case have a maximal relative error of
around $10\%$\cite{Bulla,Keil}. Flow equation approach based on
infinitesimal transformations gives its result for the sub-Ohmic
case with error $10\%$ or so\cite{3.}.  In Table I, the
generalized Shiba relation is verified for various couplings,
tunneling matrix and bath types. It shows that the Shiba's
relation is exactly satisfied for $\alpha < \alpha_c$ in both the
sub-Ohmic and Ohmic cases. Outside the range, the agreement is
still good but no longer exact. In the present paper, we consider
carefully the behavior of coherent dynamics for $\alpha <
\alpha_c$. For $s=1$, the Shiba's relation has been proved
analytically by our approach in Ref. \cite{zhe}.

The sum rule $C(t=0)=1$ is an important relation in the
equilibrium correlation function which constitutes another
significant check of our approach. The numerical solution of our
approach can yield a satisfactory result, which is also shown
obviously in Table I. As is told in the paper of Stauber\cite{3.},
the sum rule is not fulfilled by flow equation method. Independent
of the bath type and the cutoff $\omega_c$ the sum rule is
performed well in the present approach, while in the flow
equations it only yield $80\%$ or so.

Since different dynamical quantities may be associated with
different initial preparations of the system, quantum coherence
may be more or less sensitive to dissipation\cite{egger}.
Therefore, there are disparate critical values of the damping
strength for different coherence criterion. Considering the
character of the spectral function $S(\omega)$, one gives its
appropriate coherence criterion. For weak damping,
$\alpha<\alpha_c^{*}$, the function $S(\omega)$ exhibits two
inelastic peaks at finite frequency $\omega_{p} \simeq \eta
\Delta$. It can be seen that the width of peak $\gamma(\omega)$ is
less than the value of $\omega_{p}$. At the critical damping
$\alpha_c^{*}$, one has $\gamma(\omega)= \omega_{p}$, and the two
peaks merge into a single quasielastic peak centered at
$\omega=0$. For instance, in the scaling limit, for $s=1$, the
critical damping value is determined by
\begin{equation}
\frac{\omega}{\eta\Delta}-1+\frac{1}{\pi}(1+\frac{\omega}{\eta\Delta}
-\frac{\omega}{\eta\Delta}\ln\frac{\omega}{\eta\Delta})=0,
\end{equation}
and
\begin{equation}
2\alpha_c^{*}(\frac{1}{1+\frac{\omega}{\eta\Delta}})^2=1.
\end{equation}
One gets $\alpha_c^{*}=0.325$, which agrees well with
$\alpha_c^{*}=\frac{1}{3}$ or $\alpha_c^{*} \approx 0.3$ obtained
by the renormalization group numerical results \cite{Bulla}and
other conjectures\cite{Costi,lesage}. One can observe that the
"quasiparticle peak" disappears at $\alpha=\alpha_c^{*}$ in Fig. 6
(b). Physically, it indicates that the crossover between the
underdamping oscillating behavior and overdamping decaying one
occurs at $\alpha=0.325$ at zero temperature.

In Fig. 6 the renormalized spectral function $S(\omega)\Delta^{2}$
is shown for $s=0.8$ (a) and $s=1$ (b) with $\Delta/\omega_s=1$ ,
$\omega_s/\omega_c=0.1$ and various coupling strength $\alpha$.
$S(\omega)$ has a double-peak structure for $\alpha<\alpha_c^{*}$.
Only the $\omega \geq 0$ part is shown in Figs. 6(a) and 6(b). For
$\alpha \ge \alpha_c^{*}$ there is only one peak at $\omega=0$.
The critical coupling $\alpha_c^{*}$ corresponds to the crossover
from underdamping to overdamping oscillations. We obtain
$\alpha_c^{*}=0.34$ for a subohmic bath with $s=0.8$ which is in
good agreement with $\alpha_c^{*} \approx 0.3 $ obtained by the
flow equations\cite{2.}. Fig. 6(c) shows the $S(\omega)$ versus
$\omega$ relations for fixed $\alpha=0.3$ and $s=0.9$ with
$\Delta/\omega_s=1$, $5$, $10$ and $30$, respectively.

The dynamical phase diagrams are given in Fig. 7(a) ($s=1$) and
Fig. 7(b) ($s=0.8$) for $\omega_s=\omega_c$. In both figures, the
transition boundary between the incoherent relaxation and coherent
oscillation is shown in a solid line, while that of
underdamping-overdamping transition is shown in a dashed line. In
the scaling limit ($\Delta'<<1$) for $s=1$, $\alpha_c=0.5$ and
$\alpha_c^{*}=0.325$ are consistent with the known
results\cite{rmp,Costi,lesage}. Further, in the Ohmic case, the
coherent-incoherent transition coupling
$\alpha_c=\frac{1}{2}[1+\eta\Delta/\omega_c]$. With increasing
tunneling, both $\alpha_c$ and $\alpha_c^{*}$ become larger, which
presents a contrast with the tendency of the boundary obtained by
Monte Carlo simulations\cite{vol}. Moreover, note that the
critical coupling is sensitive to the bath type, especially in the
scaling limit. As displayed in Fig. 7(b), the critical couplings
$\alpha_c^{*}$ and $\alpha_c$ nicely follow power laws as
functions of the bare tunneling, $\alpha \propto \Delta^{1-s} $
for small $\Delta$ (in the case $s=0.8$). Our fit data for
$\Delta' < 0.01 $ are well consistent with the conclusion. It is
pointed out by our approach that the critical couplings in
addition to the delocalized-localized transition one are always
proportional to $\Delta^{1-s}$ for small $\Delta$, which hints
that the critical value is sensitively dependent on the structure
of bath.

\section{Summary and discussion}
The dynamics of SBM with a sub-Ohmic bath is studied by means of the
perturbation approach based on a unitary transformation. Our
approach is quite simple, whereas it correctly gives $P(t)$, $C(t)$,
Shiba's relation,  and reproduces nearly all results of previous
authors that we note. By means of our approach, the dynamical
transition point $\alpha_c$ is calculated for the first time. The
main results include: (1) the non-equilibrium correlation $P(t)$,
the susceptibility $\chi^{\prime\prime}(\omega)$ and the equilibrium
correlation $C(t)$ are analytically obtained for the general finite
$\Delta/\omega_c$ case; (2) a critical transition point ($\alpha_l$)
from delocalization to localization exists in the sub-Ohmic case;
(3) for a fixed tunnelling, as the coupling with environment
increases, a crossover from coherent to incoherent tunnelling
appears. In other words, for $\alpha>\alpha_c$, the coherent
dynamics disappear; (4) there is a critical transition point
$\alpha_{c}^{*}$ from underdamping oscillation to overdamping one in
the susceptibility $\chi^{\prime \prime}(\omega)/\omega$. When
$\alpha < \alpha_{c}^{*}$, the spectrum of susceptibility is of a
double-peak structure, while there is only one peak at $\omega=0$
for $\alpha>\alpha_{c}^{*}$; (5) the Shiba's relation and sum rules
are exactly satisfied for $\alpha \leq \alpha_c$;  (6) the dynamical
phase diagrams in both the sub-Ohmic and Ohmic cases are mapped out.
For small tunneling, all of three critical coupling $\alpha_l,
\alpha_c, \alpha_c^{*}$ are proportional to $\Delta^{1-s}$.

Our treatment is essentially a perturbation one. The transformed
hamiltonian is divided into the unperturbed part $H'_{0}$ and
perturbation $H'_{1}$ and $H'_{2}$. The unperturbed part gives the
intrinsic coherent tunnelling motion with diagonal transition of
bosons. $H'_1$ is related to the single-boson non-diagonal
transition and $H'_2$ to the multi-boson transition. In our
perturbation treatment we take into account only the single-boson
non-diagonal transition ($H'_{1}$) which plays an important role for
weak coupling and lower temperature. $H'_{2}$ with multi-boson
non-diagonal transition is neglected in our treatment. At strong
coupling and high temperature, the multi-boson non-diagonal
transition may play an important role. Since our perturbation
treatment keeps the contribution of the single-boson non-diagonal
transition and drops that of the multi-boson non-diagonal
transition, we can not make definite statement about the exact
nature of the localization-delocalization transition. Our
perturbation treatment could not be applied to study the localized
phase beyond the critical point of the sub-Ohmic case($\alpha
> \alpha_l$). Thus, the approach
can not predict the oscillation behaviors on short times in the
localized phase for strongly sub-Ohmic case. The recent NRG's
results have captured the coherent oscillation even in the localized
phase for small $s$ and demonstrated it explicitly\cite{anders}. In
principle, our approximation can be applied satisfactorily for
$\alpha \leq \alpha_c$, whereas the approximation might fail when
$\alpha \simeq \alpha_l$.

At lower temperature and weak coupling, the diagonal transition of
bosons dominates with coherent tunnelling motion. As the coupling
increases, single-boson and multi-boson non-diagonal transitions
come to prevail against the diagonal transition, which result in
decoherence. Our perturbation treatment keeps the contribution of
the single-boson non-diagonal transition and drops that of the
multi-boson non-diagonal transition. Based on the perturbation
treatment, our approach can reproduce the well-known results for the
Ohmic case, such as $\alpha_l=1$ and $\alpha_c=1/2$. Besides, Our
result also turns out that the dynamical property is not determined
only by the single energy scale $\eta \Delta$. On the one hand, a
non-scaling behavior of dynamical quantities is shown in Fig. 3(b).
On the other hand, from the results of $C(\omega)$ for $s=0.6$ and
$s=0.3$ (Figs. 5(a) and 5(b)), one can see that a pronounced
shoulder in $C(\omega)$ comes to appear with increasing coupling
which agrees with the NRG results. These complicated behaviors
result from the non-diagonal transition, which is the contribution
of $H'_{1}$ in our perturbation treatment. At the same time, Shiba's
relation and sum rule are exactly satisfied for $\alpha \leq
\alpha_c$ in both the Ohmic case and sub-Ohmic one ( even for $s \ll
1$ )which constitute important checks of the approach. In general,
this method works well for weak coupling with $\alpha \leq
\alpha_c$. Thus, this approach can be applied in the sub-Ohmic case
to study the coherent-incoherent transition.

The function $\xi_k$ plays an important role in our approach. It
eliminates the infrared catastrophe completely. In conventional
perturbation theory, from the original Hamiltonian $H$, the
dimensionless expansion parameter is $g^2_k/\omega^2_k$. Thus, the
renormalized tunneling becomes zero, and the system is always
localized because
\begin{equation}
\sum_k\frac{g_k^2}{\omega_k^2}=\int_{0}^{\omega_c}
\frac{2\alpha\omega_s^{1-s}}{\omega^{2-s}}d\omega.
\end{equation}
For Ohmic bath $s=1$  the integrand becomes $2\alpha /\omega$
which is logarithmic divergent in the infrared limit. There also
exists a low-energy divergence ($\frac{1}{\omega^{2-s}}$) for a
sub-Ohmic bath $s<1$.  On the contrary, for the coupling in
transformed Hamiltonian, $H'_1$, the expansion parameter is
$g^2_k\xi^2_k/\omega^2_k\sim
2\alpha\omega^{s}\omega_s^{1-s}/(\omega+\eta\Delta)^2$, which is
finite in the infrared limit for $s>0$. Therefore, from the
viewpoint of mathematics, the introduction of $\xi_k$ gets rid of
the divergence. Physically, the disappearance of divergence arises
due to the energy scale separation by $\xi_k$ with a special
treatment of the low frequency, nonadiabatic modes.

The low frequency behavior of the spectrum density function in SBM
model determines the long time behavior of TSS. All quantum
dynamics properties are very sensitive to the low energy part in
spectral structure, especially in the sub-Ohmic case. On the other
hand, in the high frequency limit($\omega_k >> \Delta$), bath
modes follow instantaneously to the tunneling motions, namely, the
displacement of boson due to the coupling to subsystem is large,
whereas near the low frequency limit($\omega_k << \Delta$),
nonadiabatic modes couple weakly to the subsystem with coupling
strength $g_k\frac{\omega_k}{\eta\Delta}$, its displacement is
small. Thus, while all boson modes are treated by the function
$\xi_k$, their different contribution to the dressed TSS has been
distinguished with respect to the scale of boson energy.
Therefore, even if our approach is, in principle, a perturbation
one, those reasons above give the intrinsic base of our present
results, such as $P(t)$, $C(t)$, the dynamical transition points
$\alpha_c$ and $\alpha_c^{*}$ and exact Shiba's relation. When $s
\rightarrow 1$, our results agree well with exact or known ones.
Besides, our approach can be straightforwardly extended to other
more complicated coupling systems.

\vskip 0.5cm

{\noindent {\large {\bf Acknowledgement}}}

We would like to thank Dr. Ninghua Tong for useful discussions and
providing the data of numerical renormalization group. This work
is supported by the China National Natural Science Foundation
(Nos.10474062, 10547126 and 90503007). The work was partially
supported by the Shanghai Jiao Tong University Youth Foundation.
We would like to thank the High-performance Computing Center in
Shanghai Jiao Tong University.

\rm 

\newpage
\begin{center}
{\Large \bf Figure Captions }
\end{center}

\vskip 0.5cm

\baselineskip 16pt

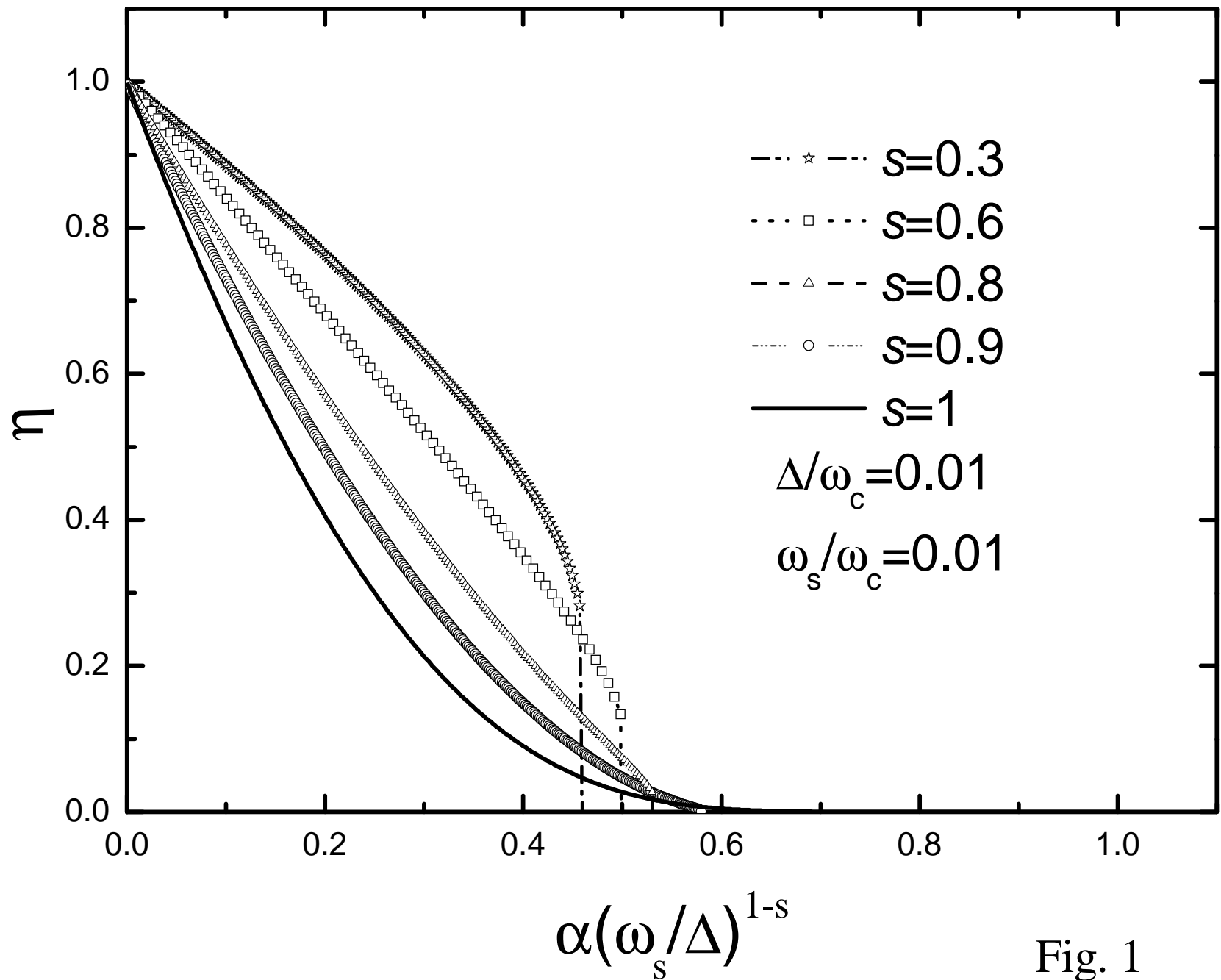
{\bf Fig.1}~~~$\eta$ as a function of the dimensionless coupling
strength $\alpha (\frac{\omega_s}{\Delta})^{1-s}$ for different
bath types.

\vskip 0.5cm

{\bf Fig.2}~~~(a)The delocalized-localized transition point
$\alpha_l$ as functions of $s$. (b) $\Delta$ dependence of the
critical coupling $\alpha_l$ for various bath types $s$. The NRG
data are also shown for comparison. NRG parameters here are
$\Lambda = 2$, $Nb = 8$, and $Ns = 100$.

\vskip 0.5cm

{\bf Fig.3}~~~(a)The time evolution ($\Delta_r t=\eta\Delta t$)of
the non-equilibrium correlation $P(t)$ for $\alpha
\omega'^{1-s}_s=0.1$ and $\Delta/\omega_s=10$ with different bath
types $s$. (b) $P(t)$ for $s=0.9$ and $\alpha=0.2$ with various
values of $\Delta$. The inset shows $P(t)$ for $\alpha=0.1$ with
different $\Delta$ in the Ohmic case.

\vskip 0.5cm

{\bf Fig.4}~~~The time evolution of $P(t)$ and $C(t)$ for $s=0.8$
and $\Delta/\omega_s=5$ with different couplings$\alpha=0.1,0.3$,
and $0.5$, respectively. Note that $\alpha_c=0.53718$.

\vskip 0.5cm

{\bf Fig.5}~~~The equilibrium correlation function $C(\omega)$ for
(a) $s=0.6$, $\Delta'=0.1$ and $\alpha
\omega'^{1-s}_s=0.05,0.08,0.12$, and $0.13$; (b) $s=0.3$ with
various values of $\alpha$, and $\Delta'=0.1$
($\omega_s=\omega_c=1$). we find $\alpha_c=0.1271$ for $s=0.6$ and
$\alpha_c=0.0327$ for $s=0.3$.

\vskip 0.5cm

{\bf Fig.6}~~~The renormalized spectral function
$S(\omega)\Delta^{2}$ as a function of $\omega$ for $s=0.8$ (a)
and $s=1$ (b) with $\Delta/\omega_s=1$ , $\omega_s/\omega_c=0.1$
and different $\alpha$.   (c) $S(\omega)\Delta^{2}$ as a function
of $\omega$ for $s=0.9, \alpha=0.3$ with $\Delta/\omega_s=1,5,10$,
and $30$, respectively.

\vskip 0.5cm

{\bf Fig.7}~~~The dynamical phase diagrams for $s=1$(a) and
$s=0.8$(b) with $\omega_s=\omega_c$. The solid lines are the
transition boundaries between the coherent oscillation and
incoherent decay. The dashed lines stand for the boundaries
between the underdamping oscillation and overdamping one in
coherent region. The dotted lines in fig6. (b) are fits to the
$\alpha \propto \Delta^{1-s}$ law using the $\Delta <10^{-2}$
points only.

\newpage
\begin{center}
{\Large \bf Table }
\end{center}

\vskip 0.5cm

\baselineskip 16pt

\begin{tabular}{ccccccccc}
\hline \hline s& $\omega_s'$ & $\frac{\Delta}{\omega_s}$ &
$\alpha$& $ \chi_0 $ &
$\frac{C(\omega)}{J(\omega)}|_{\omega\rightarrow 0} $ & R &
$C(t=0)$ \\ \hline
1& 1         & 0.01 & 0.1&93.275771 &34801.4793&1&1&\\
1& 1         & 0.05 & 0.1&15.588677 &972.0275&1&1&\\
1& 1         & 0.1  & 0.1&7.211567  &208.0268&1&1&\\
1& 1         & 0.2  & 0.1&3.336971  &44.54151&1&1&\\
1& 1         & 0.1  & 0.3&21.202413 &1798.1694&1&1&\\
1& 1         & 0.1  & 0.4&54.393368 &11834.5544&1&1&\\
1& 1         & 0.2  & 0.5&65.0612653&16931.8733&1&1&\\
0.9& 1         & 0.1  & 0.1&8.012265&256.7861&1&1&\\
0.9& 1         & 0.05 & 0.1&18.18289&1322.467&1&1&\\
0.9& 1         & 0.2  & 0.15&4.450706&79.23528&1&1&\\
0.8& 1         & 0.05  & 0.1&23.81572&2268.770&1&1&\\
0.8& 1         & 0.1  & 0.1&9.474312&359.0517&1&1&\\
0.8& 0.1       & 1    & 0.1&7.233729&209.3079&1&1&\\
0.6& 1         & 0.1  & 0.01&5.496693&120.8546&1&1&\\
0.5& 1         & 0.2  & 0.05&4.282478&73.35843&1&1&\\
0.5& 0.1       & 1    & 0.1&8.26555&273.2771&1&1&\\
 \hline\hline
\end{tabular}

\newpage
\begin{center}
{\Large \bf Table Captions }
\end{center}

\vskip 0.5cm

\baselineskip 16pt

{\bf TABLE I}: Representative results from the numerical solution
with parameters chosen by the spectral density
$J(\omega)=2\alpha\omega_s^{1-s}\omega^{s}\Theta(\omega_c-\omega)$
and with the controlling precision $10^{-5}$ for iteration.
$R\equiv[\lim_{\omega \to 0}C(\omega)/J(\omega)]/(2\chi_0)^2$.The
numeric error for the Shiba relation and sum rule is at least less
than $10^{-6}$ and can be improved by increasing the accuracy of
numerical calculations.

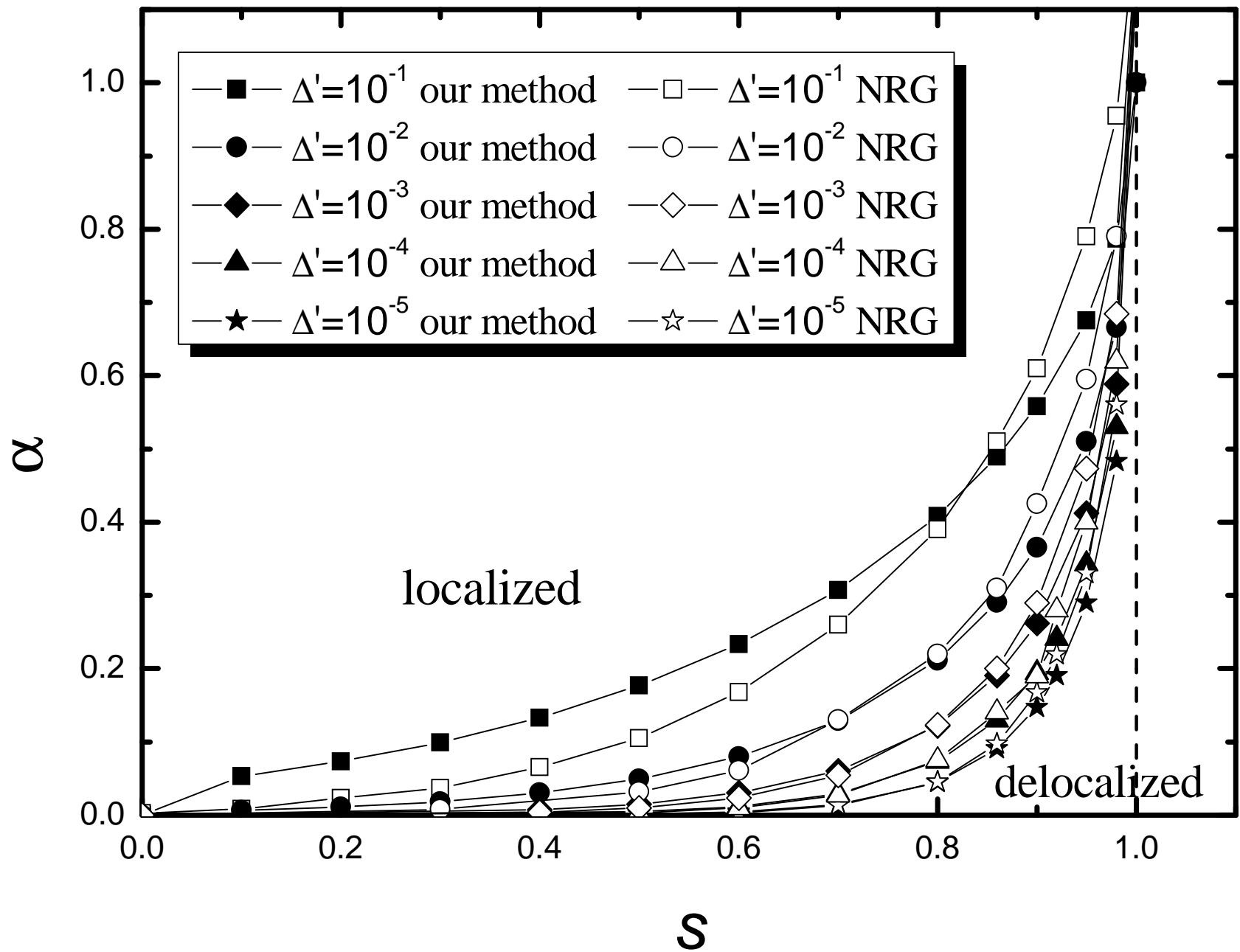
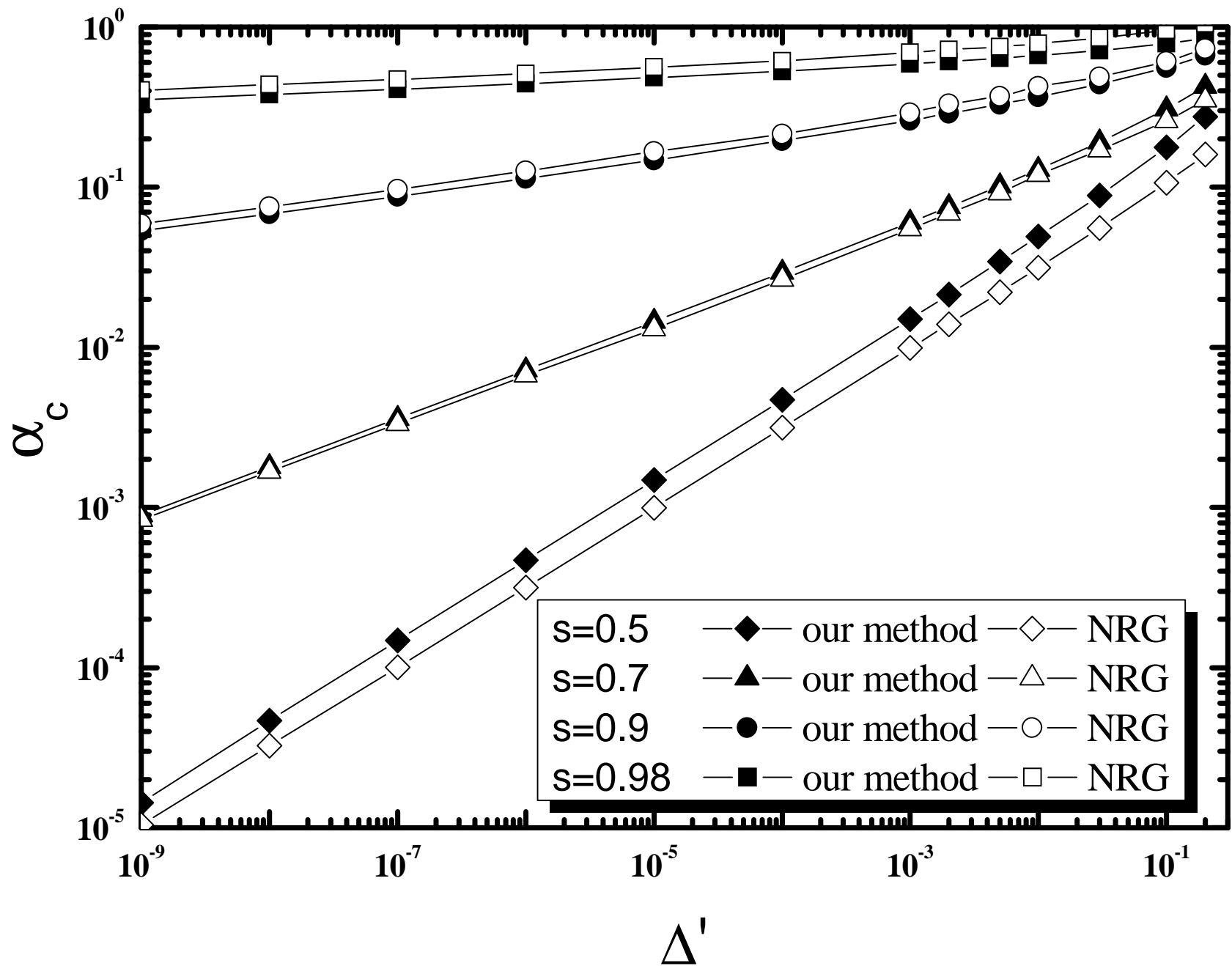
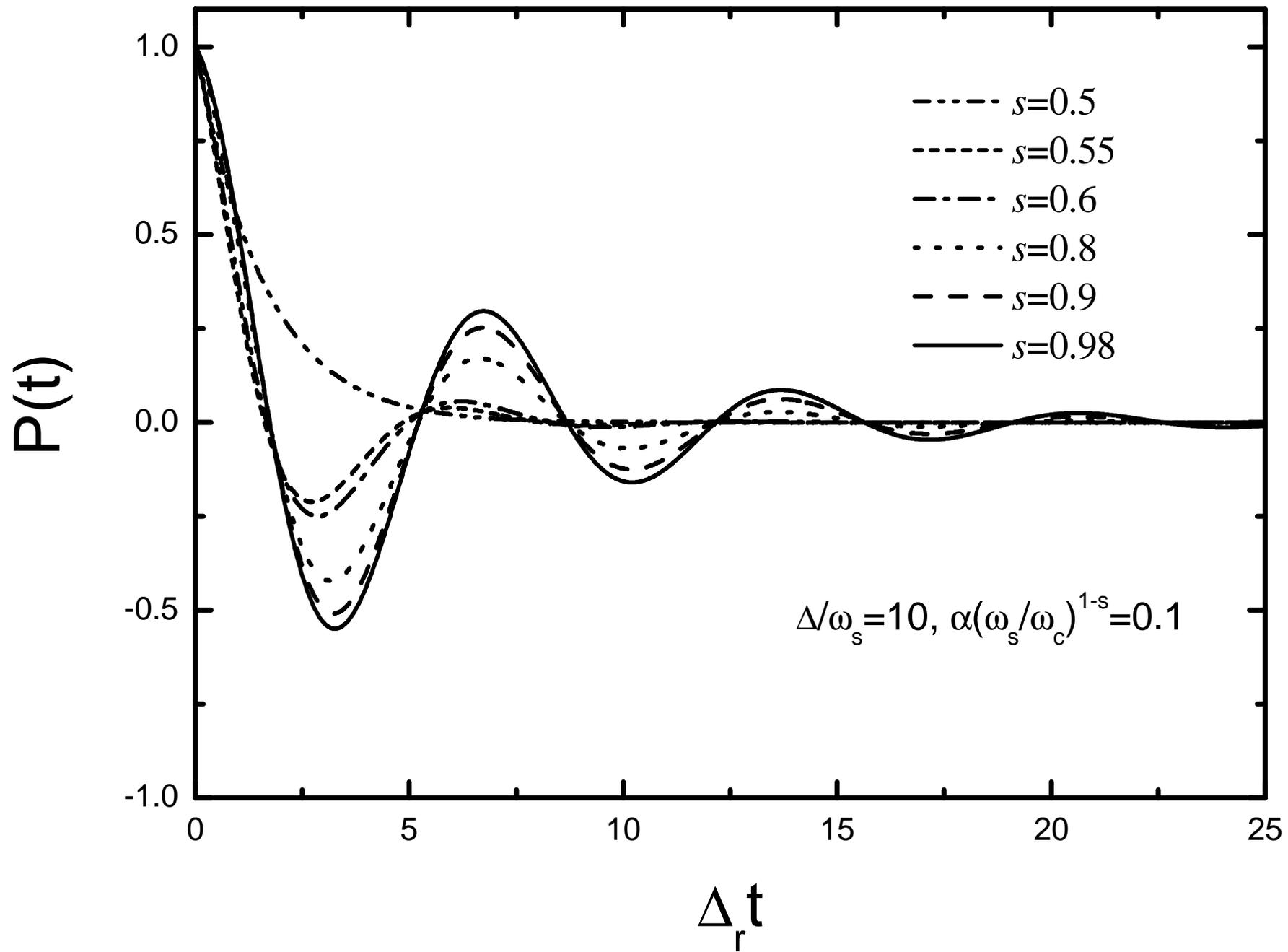
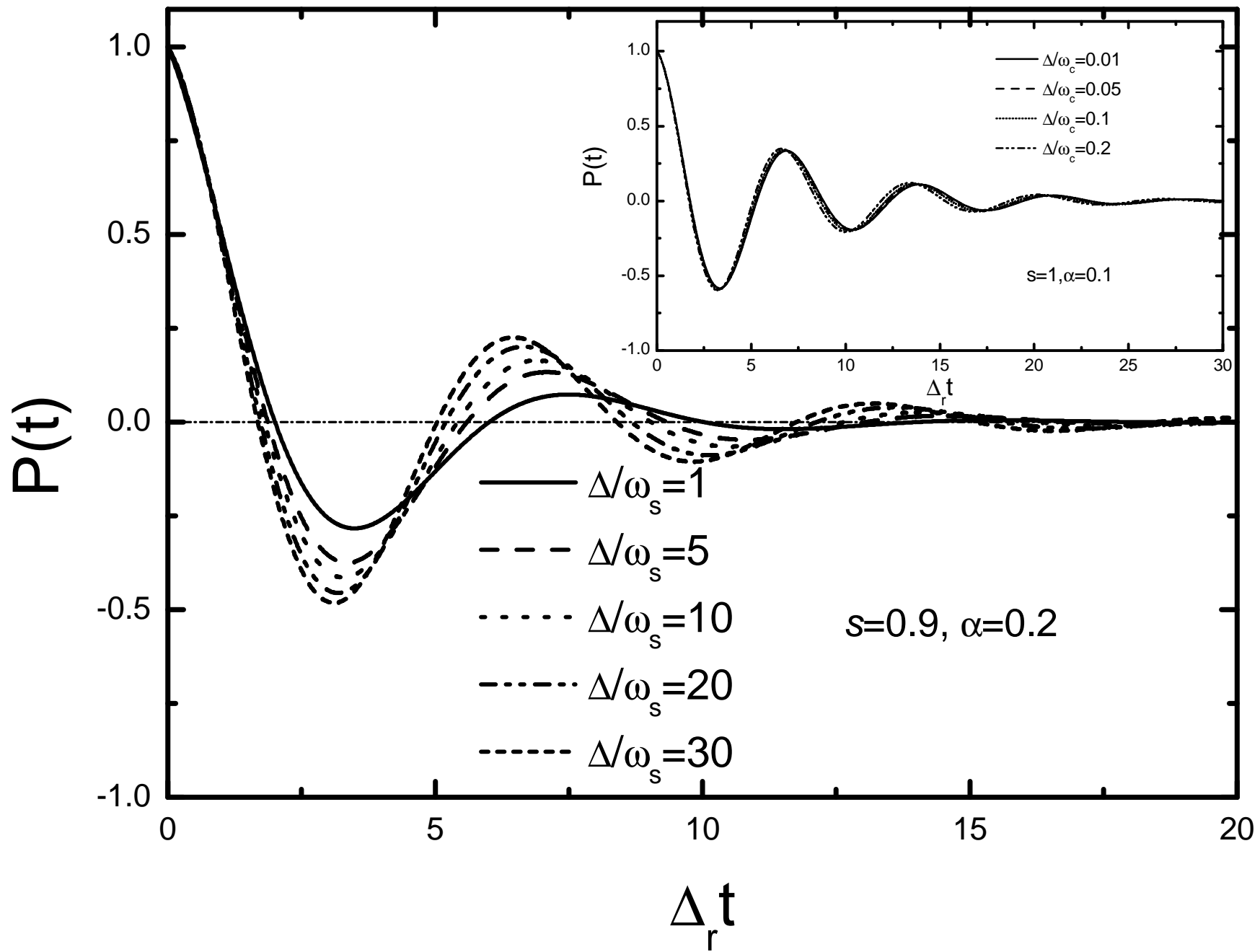
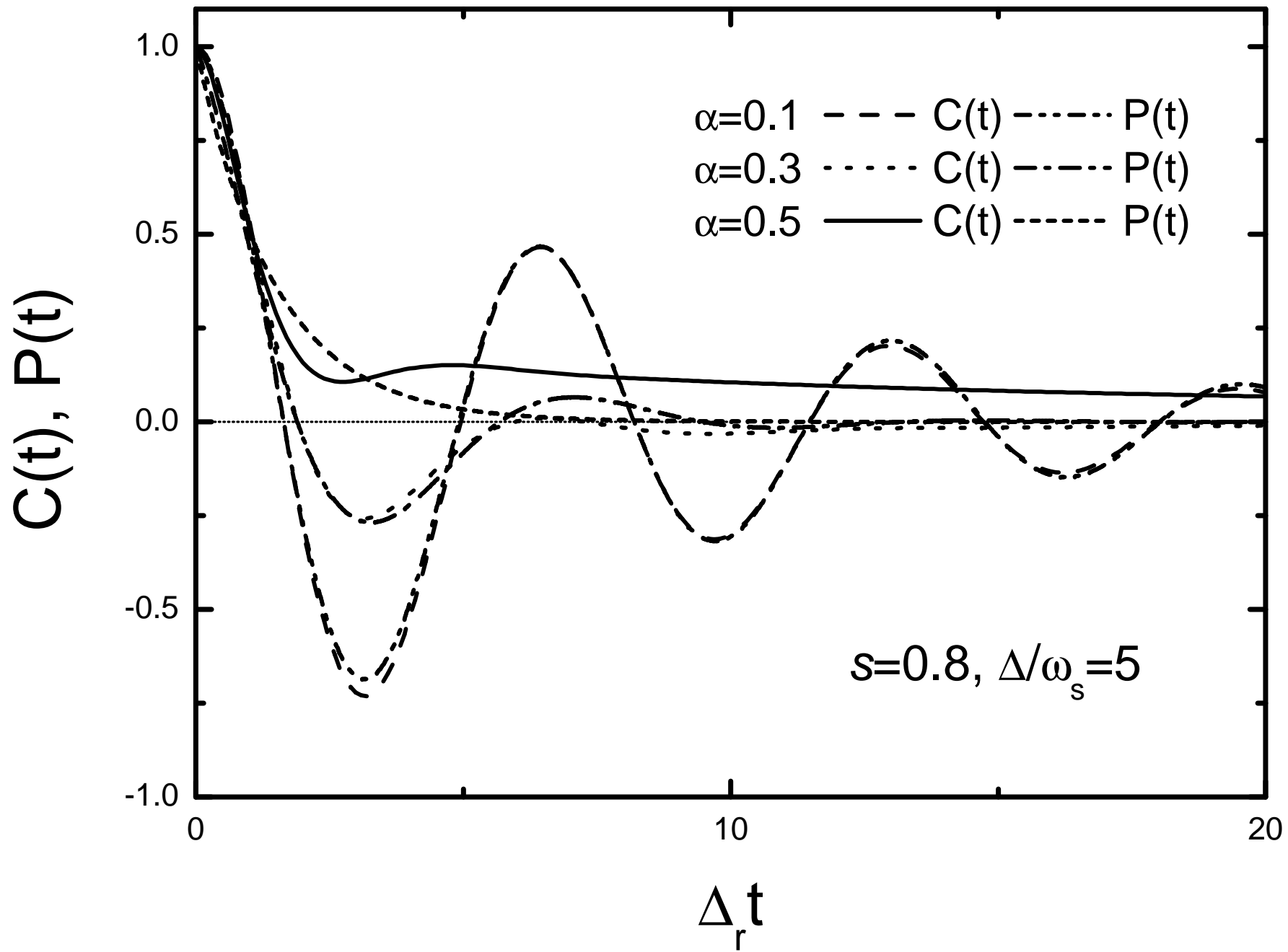
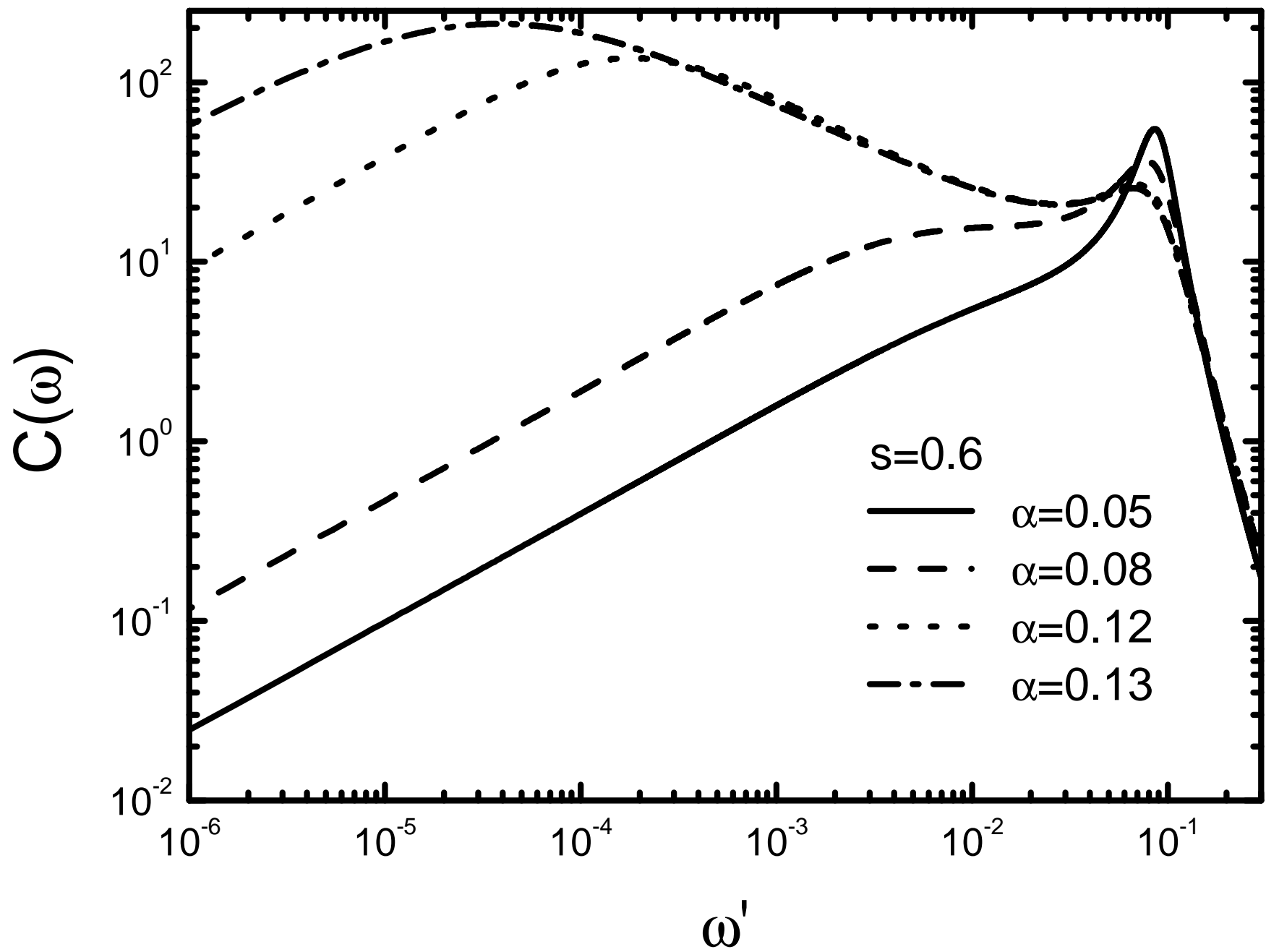
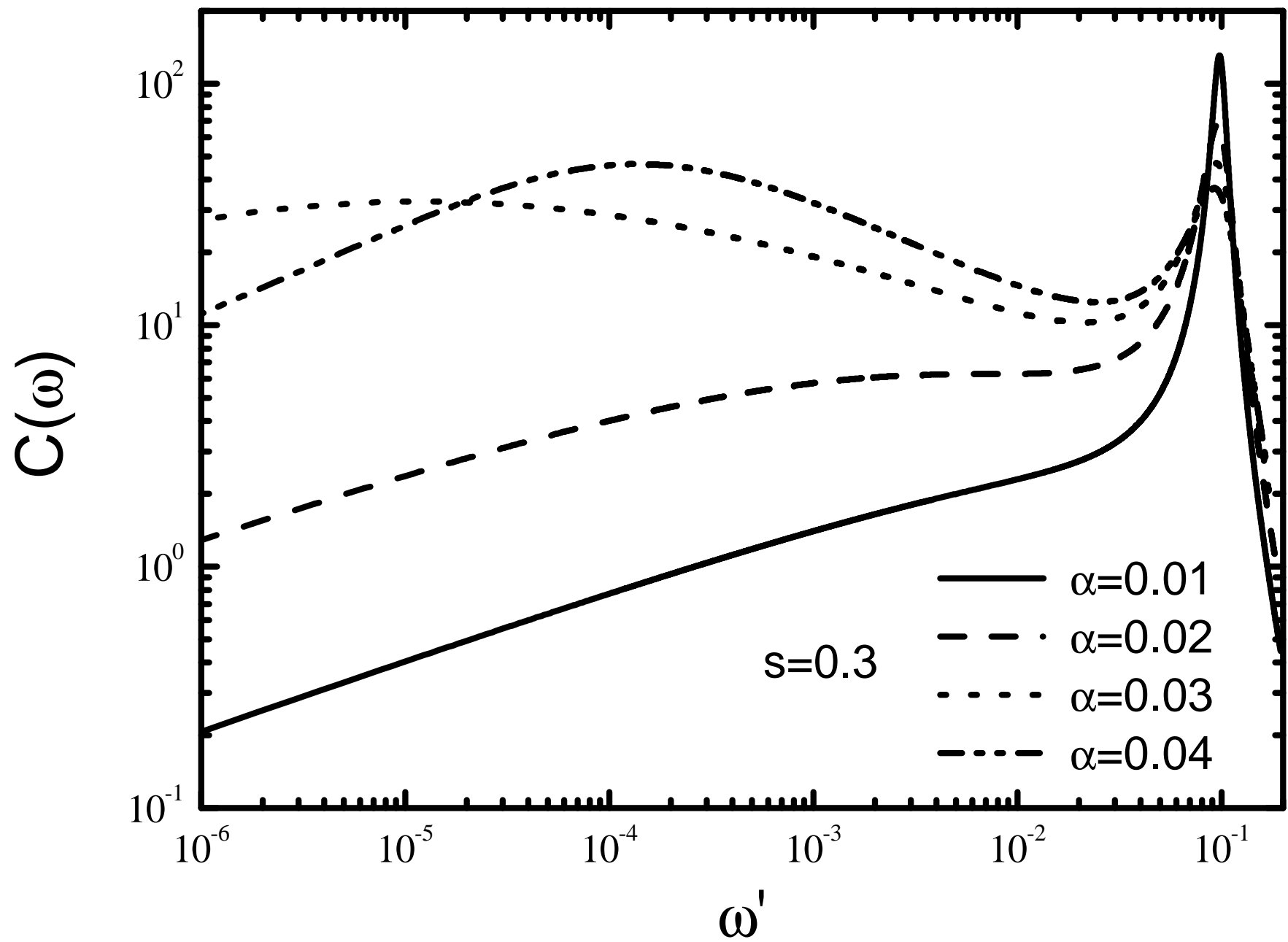
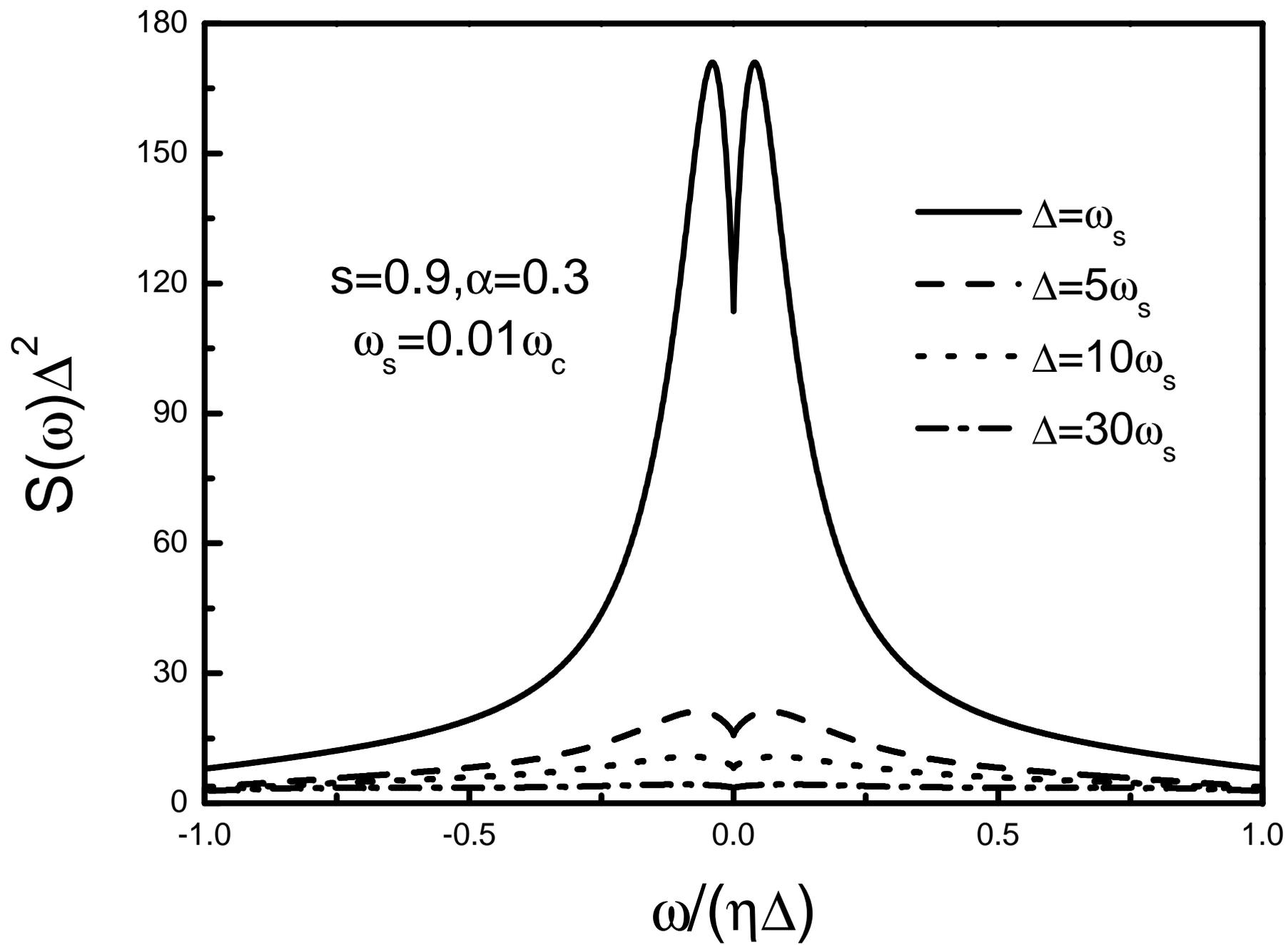
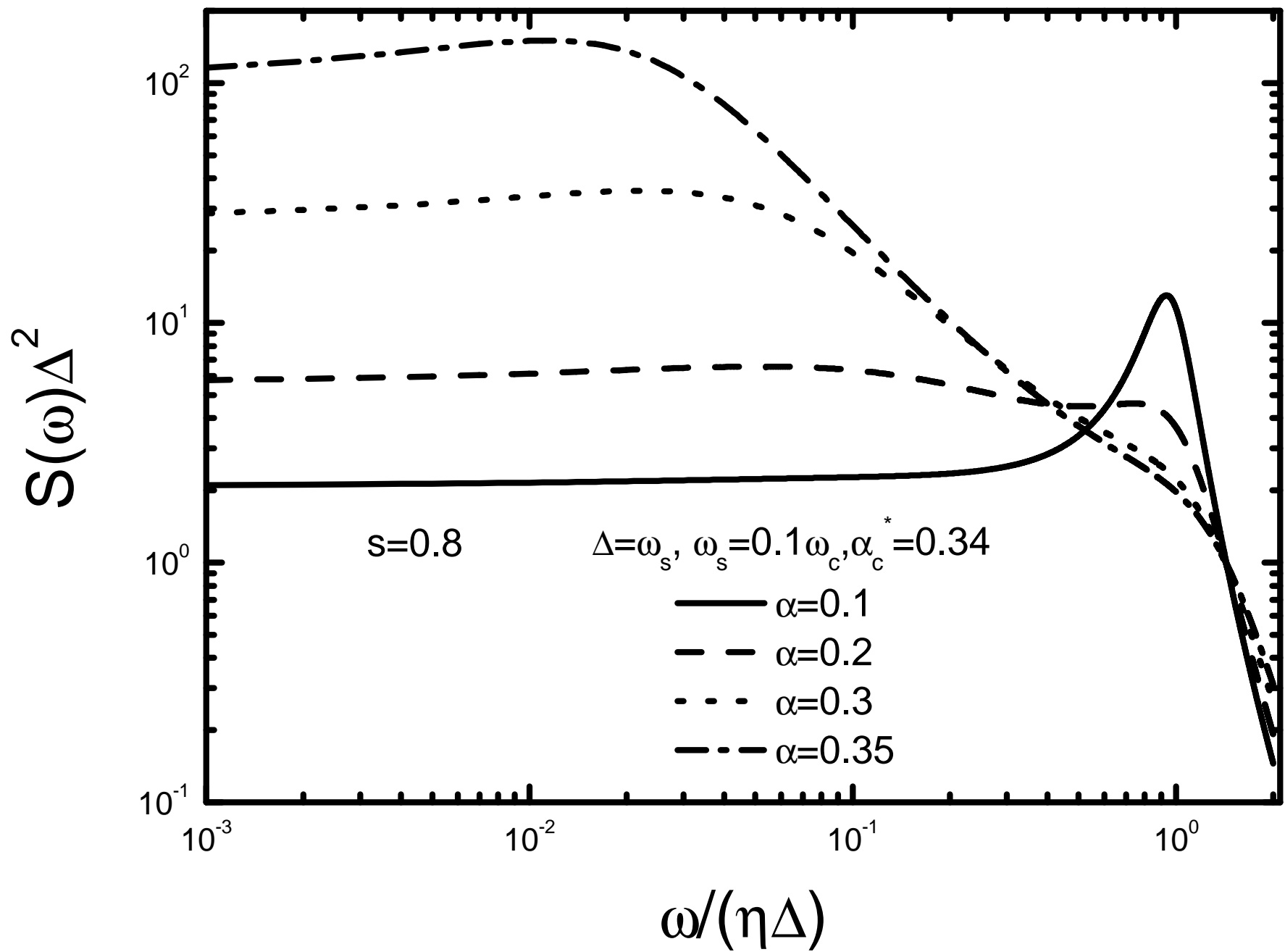
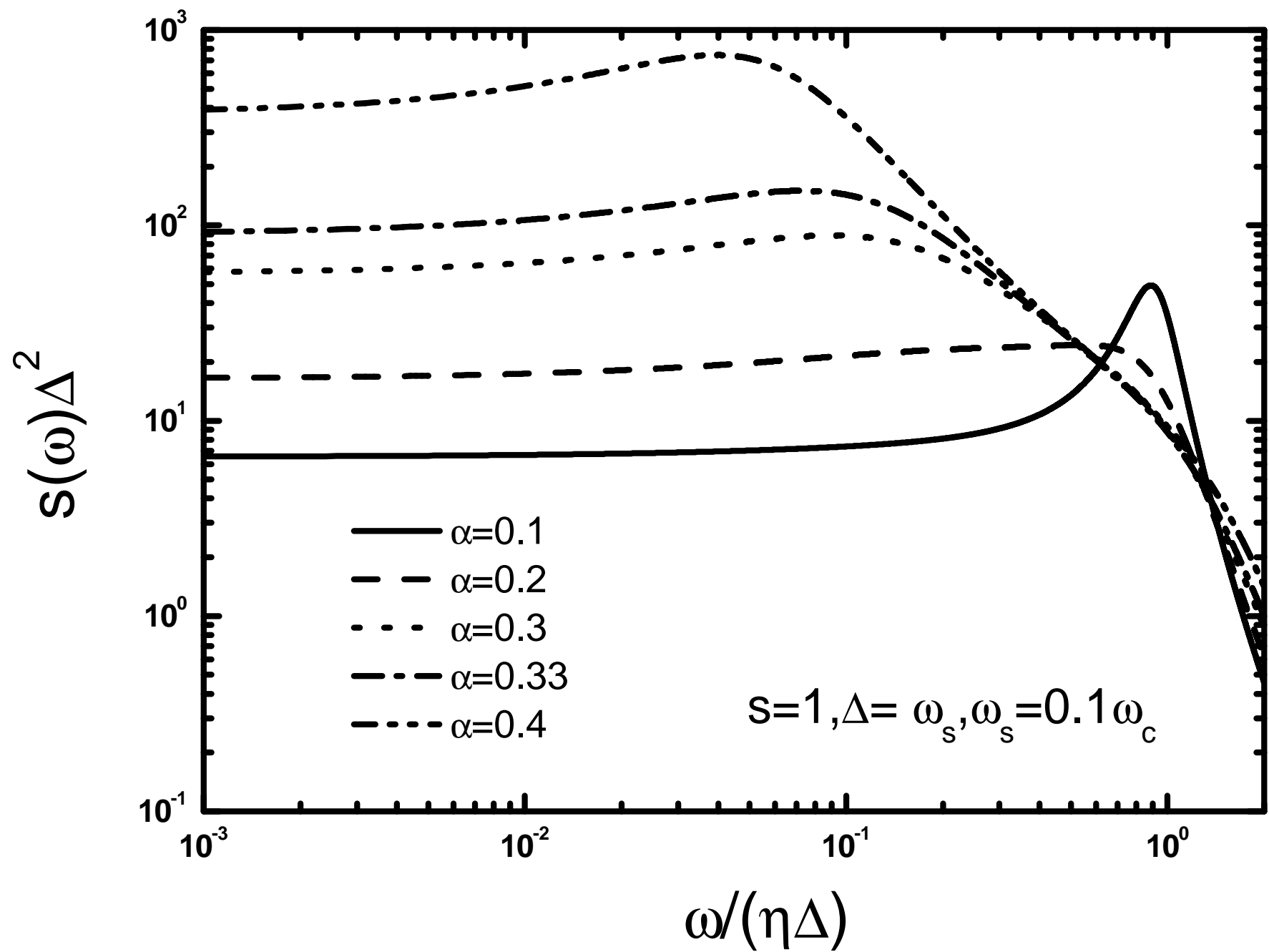
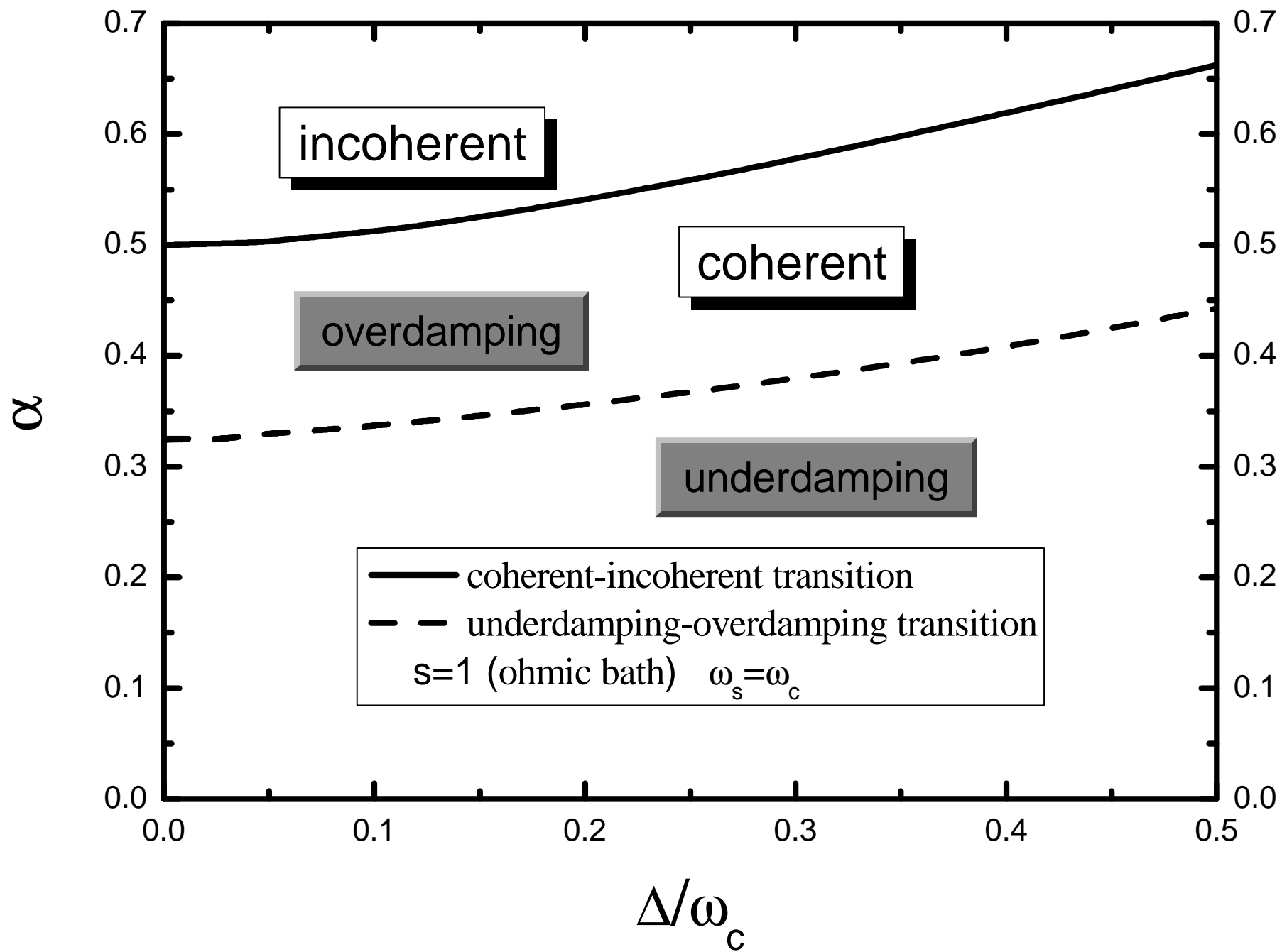
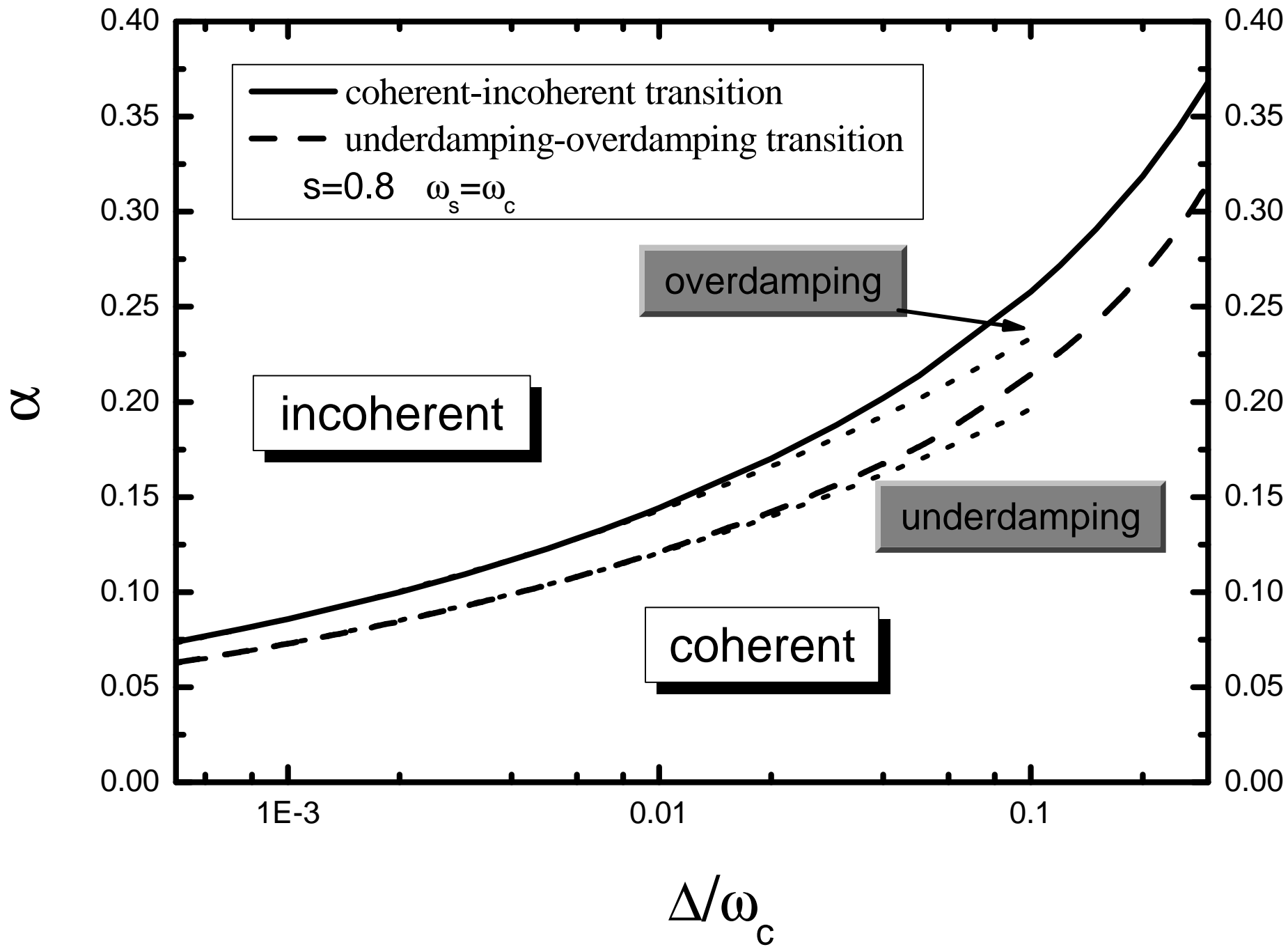
\end{document}